\documentclass[12pt,a4paper]{article}
\usepackage[doublespacing]{setspace}
\usepackage[latin1]{inputenc}
\usepackage{amsmath}
\usepackage{amsfonts}
\usepackage{amssymb}

\usepackage[english]{babel}
\usepackage{amsmath,amsthm}
\usepackage{amsfonts}
\usepackage{appendix}

\usepackage{natbib}

\usepackage{graphicx,pict2e}
\usepackage{rotating}
\usepackage{algorithm}
\floatname{algorithm}{Algorithm}

\newtheorem{prop}{Proposition}
\newtheorem{lemma}{Lemma}
\newtheorem*{lemproof}{Proof}

\numberwithin{equation}{section}

\title{Preserving the distribution function in surveys in case of imputation for zero inflated data}
\date{\today}

\author{Brigitte Gelein\thanks{Univ Rennes, ENSAI, CNRS, IRMAR - UMR 6625, F-35000 Rennes, France} and Guillaume Chauvet\thanks{Univ Rennes, ENSAI, CNRS, IRMAR - UMR 6625, F-35000 Rennes, France} }

\begin{document}

\maketitle

\begin{abstract}
\noindent Item non-response in surveys is usually handled by single imputation, whose main objective is to reduce the non-response bias. Imputation methods need to be adapted to the study variable. For instance, in business surveys, the interest variables often contain a large number of zeros. Motivated by a mixture regression model, we propose two imputation procedures for such data and study their statistical properties. We show that these procedures preserve the distribution function if the imputation model is well specified. The results of a simulation study illustrate the good performance of the proposed methods in terms of bias and mean square error.
\end{abstract}

\noindent{\small{{\it Keywords:} balanced imputation, imputation model, item non response, mixture model, regression imputation.}} \\

\normalsize

\section{Introduction}

\noindent Item non-response may affect the quality of the estimates when the respondents and the non-respondents exhibit different characteristics with respect to the variables of interest. Item non-response in surveys is usually handled by single imputation, whose main objective is to reduce the non-response bias. Two approaches are commonly used in sample surveys to motivate imputation. Under the non-response model approach (NM), the response mechanism is explicitly modeled, whereas under the imputation model approach (IM), the variable under study is explicitly modeled. \\

\noindent Single imputation consists of replacing a missing value with an artificial one. It leads to a single imputed data set, constructed so that it is possible to apply complete data estimation procedures for obtaining point estimates. The response indicators are therefore not required. On the other hand, multiple imputation methods \citep{rub:87,lit:rub:87} consist in building $M>1$ imputed datasets, and in estimating the parameters under study for each of them. The $M$ analyses are then combined for inference. Multiple imputation has been extensively studied in the literature, some recent references include \citet{iac:por:07}, \citet{whi:car:10} and \citet{tem:kow:fil:11}. However, multiple imputation is not commonly used in sample surveys. Under the NM approach, multiple imputation needs to be proper for valid inference. Some sufficient conditions are given in \cite{rub:87}, pp. 118-119, but they are usually difficult to check for complex sampling designs, see \citet{bin:sun:96}, \citet{fay:92}, \citet{fay:96} and \citet{nie:03}. Also, under the IM approach, the multiple imputation variance estimator does not track the variance correctly, and can be considerably biased, see \citet{kot:95}, \citet{kim:bri:ful:kal:06}, \citet{bri:kal:kim:06} and \citet{bea:haz:boc:11}. Therefore, we focus in this paper on single imputation methods. \\

\noindent The Imputation Model (IM) approach is of common use to treat item non-response in surveys. The imputation methods are then motivated by a modeling of the relationship between the variable of interest and the available auxiliary variables. Both the imputation model and the imputation methods need to be adapted to the study variable. For instance, in business surveys, the interest variables often contain a large number of zeros. In the Capital Expenditure Survey conducted at Statistics Canada, approximately 70\% of businesses reported a value of zero to Capital Machinery and 50\% reported a value of zero to Capital Construction \citep{haz:cha:nam:14}. In case of some interest variable containing a large amount of zeroes, \citet{haz:cha:nam:14} propose imputation methods based on a mixture regression model. They prove that these methods lead to doubly robust estimators of the population mean, i.e. the imputed estimator of the mean is consistent whether the interest variable or the non-response mechanism is adequately modeled. However, these methods are not appropriate when estimating more complex parameters such as the population distribution function. \\

\noindent In this work, we propose an imputation which enables to preserve the distribution function for zero inflated data. This is an important practical property if the data users are not only interested in estimating means or totals, but also parameters related to the distribution of the imputed variable, e.g. the Gini coefficient. We use the IM approach, without explicit assumptions on the non-response mechanism for the interest variable. We propose a random imputation method which leads to a $\sqrt{n}$-consistent estimator of the total, and to a mean-square consistent estimator of the distribution function. \\

\noindent As recalled in \citet{haz:cha:nam:14}, random imputation methods suffer from an additional variability due to the imputation variance. Three approaches have been proposed in survey sampling to reduce this variance. Fractional imputation is somewhat similar to multiple imputation, and consists in replacing some missing value with $M$ imputed values to which some weights are given \citep{kal:kis:81,kal:kis:84,fay:96,kim:ful:04,ful:kim:05}. The imputation variance decreases as $M$ increases. The second approach consists in using some standard imputation mechanism, and in modifying the imputed values in order to suppress the imputation variance \citep{che:rao:sit:00}. Finally, the third approach consists of directly imputing artificial values in such a way that the imputation variance is eliminated \citep{kal:kis:81,kal:kis:84,dev:06,cha:dev:haz:11,cha:haz:12,has:til:14,cha:cha:haz:sal:sol:18}. This last approach is of particular interest because it leads to a single imputed dataset, which is attractive from a data user's perspective, and it does not require any modification of the imputed values. \\

\noindent In this paper, we propose a balanced version of our imputation method, which enables to greatly reduce the imputation variance. It consists of randomly generating the imputed values while satisfying appropriate balancing constraints, by using an adaptation of the Cube algorithm \citep{dev:til:04,cha:dev:haz:11}. Our simulation results prove that the balanced imputation method succeeds in preserving the distribution function of the imputed variable, with large variance reductions as compared to the proposed non-balanced imputation method. In order to produce confidence intervals for the estimated parameters with appropriate coverage, we also propose variance estimators adapted from the linearization variance estimators proposed by \cite{kim:rao:09}. Our simulation results indicate that these estimators perform well, both in terms of relative bias and of coverage rate. \\

\noindent The paper is organized as follows. In Section \ref{sec2}, we describe the theoretical set-up and the notation used in the paper. In Section \ref{sec3}, we briefly recall the two imputation procedures proposed by \citet{haz:cha:nam:14}, and introduce our two proposed imputation methods. In Section \ref{sec4}, we prove that the proposed random imputation procedure yields a consistent estimator of the total and of the distribution function. Variance estimation for the imputed estimator of the total is discussed in Section \ref{sec5}. The results of a simulation study comparing the four procedures and evaluating the proposed variance estimator are presented in Section \ref{sec6}. An application of the proposed methodology on data modelled in the Monthly Retail Trade Survey is presented in Section \ref{sec:appli}. We conclude in Section \ref{sec7}. All the proofs are given in the Appendix. Some additional simulation results are available in the Supplementary Material.

\section{Theoretical set-up} \label{sec2}

\noindent We are interested in some finite population $U$ of size $N$, with some quantitative variable of interest $y$ taking the value $y_i$ for unit $i \in U$. We note $y_U=(y_{1}, \ldots, y_{N})^\top$ for the vector of values for the variable $y$. We are interested in estimating the total $t_y=\sum_{i \in U} y_i$, and the finite population distribution function
    \begin{eqnarray} \label{sec2:eq1}
      F_N(t) & = & \frac{1}{N} \sum_{i \in U} 1(y_i \leq t)
    \end{eqnarray}
where $1(\cdot)$ is the indicator function.\\

\noindent A sample $s$ of size $n$ is selected according to a sampling design $p(.)$, with $\pi_i$ the first-order inclusion probability in the sample for unit $i$. We suppose that $\pi_i>0$ for any unit $i \in U$, and we note $d_i=\pi_i^{-1}$ the design weight. We note $\delta_U=(\delta_1, \ldots, \delta_N)^\top$ for the vector of sample membership indicators. In case of full response, a complete data estimator of $t_y$ is the expansion estimator or Horvitz-Thompson~(1952) \nocite{hor:tho:52} estimator
    \begin{eqnarray} \label{sec2:eq2}
      \hat{t}_{y\pi} & = & \sum_{i \in s} d_i y_i.
    \end{eqnarray}
This estimator is design-unbiased for $t_y$, in the sense that $E_p(\hat{t}_{y\pi})=t_y$ with $E_p$ the expectation under the sampling design $p(.)$, conditionally on $y_U$. We note $V_p$ the variance under the sampling design $p(.)$. Concerning the population distribution function $F_N$, plugging into (\ref{sec2:eq1}) the expansion estimators of the involved totals yields the plug-in estimator
    \begin{eqnarray} \label{sec2:eq3}
      \hat{F}_N(t) = \frac{1}{\hat{N}_{\pi}} \sum_{i \in s} d_i 1(y_i \leq t) & \textrm{ with } & \hat{N}_{\pi}=\sum_{i \in s} d_i.
    \end{eqnarray}
Under some mild assumptions on the variable of interest and the sampling design \citep[see][]{dev:99,car:cha:gog:lab:10}, $\hat{F}_N(t)$ is approximately unbiased and mean-square consistent for $F_N(t)$. \\

\noindent We now turn to the case when the variable of interest $y$ is subject to missingness. Let $r_{i}$ be the response indicator, such that $r_{i}=1$ if unit $i$ responded to item $y$, and $r_{i}=0$ otherwise. Let $p_i$ be the response probability of some unit $i$. We note $r_U=(r_{1}, \ldots, r_{N})^\top$ for the vector of response indicators. We assume that each unit responds independently of one another. Let $E_q$ and $V_q$ denote the expectation and variance under the non-response mechanism, conditionally on the vector $y_U$ of population values and on the vector ${\delta}_U$ of sample membership indicators. An imputation mechanism is used to replace some missing value $y_i$ by an artificial value $y_{i}^*$. An imputed estimator for $t_y$ based on observed and imputed values is
    \begin{eqnarray} \label{sec2:eq4}
      \hat{t}_{yI} & = & \sum_{i \in s} d_i r_i y_i + \sum_{i \in s} d_i (1-r_i) y_i^*.
    \end{eqnarray}
Similarly, an imputed estimator of the distribution function based on observed and imputed values is
    \begin{eqnarray} \label{sec2:eq5}
      \hat{F}_I(t) & = & \frac{1}{\hat{N}_{\pi}} \left\{\sum_{i \in s} d_i r_i 1(y_i \leq t) + \sum_{i \in s} d_i (1-r_i) 1(y_i^* \leq t)\right\}.
    \end{eqnarray}
In comparison with the estimators obtained in (\ref{sec2:eq2}) and (\ref{sec2:eq3}) with complete data, there are two additional random mechanisms involved in the estimators given in (\ref{sec2:eq4}) and (\ref{sec2:eq5}). First, the non-response mechanism leads to observe the values of $y$ for a part of $s$ only. Then, the imputation mechanism is used to replace missing $y_i$'s with artificial values. \\

\noindent The imputation mechanism is motivated by an imputation model, which is a set of assumptions on the variable $y$ subject to missingness. In this paper, we are interested in some quantitative variable of interest for which the imputation model may be described as a mixture regression model, see equation (\ref{imput:model}) below. The proposed imputation methods are therefore not suitable for categorical variables. An extension of the proposed imputation methods to cover such variables is discussed in our conclusion. \\

\noindent In the context of a zero-inflated variable of interest, the mixture regression model introduced in \citet{haz:cha:nam:14} is as follows:
    \begin{eqnarray} \label{imput:model}
      y_i & = & \eta_i \left\{z_i^{\top} \beta + \sqrt{v_i} \epsilon_i \right\},
    \end{eqnarray}
where the $\eta_i$'s are independent Bernoulli random variables equal to $1$ with probability $\phi_i$, and equal to $0$ otherwise; the $\epsilon_i$'s are independent and identically distributed random variables of mean $0$, variance $\sigma^2$, and with a common distribution function $F_{\epsilon}$; the parameters $\beta$ and $\sigma$ are unknown, and $v_i$ is a known constant. The vector of auxiliary variables $z_i$ is assumed to be known on the whole sample including non-respondents. To sum up, according to the imputation model (\ref{imput:model}) the variable $y_i$ follows a regression model with a probability $\phi_i$, and is equal to $0$ otherwise. \\

\noindent Note that no assumptions are made on some specific distribution for the residuals $\epsilon_i$. We only suppose that they share a common distribution function $F_{\epsilon}$. Let $E_m$ et $V_m$ denote respectively the expectation and variance under the imputation model. We suppose that the sampling design is non-informative, in the sample that the vector $\delta_U$ of sample membership indicators is independent of $\epsilon_U = (\epsilon_1,\ldots,\epsilon_N)^{\top}$ and $\eta_U=(\eta_1,\ldots,\eta_N)^{\top}$, conditionally on a set of design variables. \\

\noindent In practice, the $\phi_i$'s are unknown and need to be estimated. We assume that they may be parametrically modeled as
    \begin{eqnarray} \label{model_phi_k}
    \phi_i & = & f(u_i,\gamma)
    \end{eqnarray}
where $f$ is a known function, $u_i$ is a vector of variables recorded for all sampled units, and $\gamma$ is an unknown parameter. An estimator of $\phi_i$ is
    \begin{eqnarray} \label{estim_phi_i}
    \hat{\phi}_i  & = & f(u_i,\hat{\gamma}_r)
    \end{eqnarray}
with $\hat{\gamma}_r$ an estimator of $\gamma$ computed on the responding units. We assume that $\eta_i$ and $\epsilon_i$ are independent, conditionally on the vectors $z_i$ and $u_i$. \\

\noindent In this paper, we use the Imputation Model (IM) approach where the inference is made with respect to the imputation model, the sampling design, the response mechanism and the imputation mechanism. This does not require an explicit modeling of the non-response mechanism unlike the Non-response Model approach \citep{haz:09}, but we assume that the data are missing at random, which means that model (\ref{imput:model}) holds for both the respondents and the non-respondents. We note $E_I$ and $V_I$ the expectation and variance under the imputation mechanism, conditionally on the vectors $y_U$, ${\delta}_U$ and $r_U$.

\section{Imputation methods} \label{sec3}

\noindent In this Section, we first briefly recall in Sections \ref{ssec31} and \ref{ssec32} the random imputation methods proposed by \citet{haz:cha:nam:14} for zero-inflated data. We then introduce the new methods that we propose in Sections \ref{ssec33} and \ref{ssec34}.

\subsection{Haziza-Nambeu-Chauvet random imputation} \label{ssec31}

\noindent A first proposal of \citet{haz:cha:nam:14} is to use the imputation mechanism
    \begin{eqnarray} \label{imput:mecha:1}
      y_i^* & = & \eta_i^* \left\{z_i^{\top} \hat{B}_r \right\},
    \end{eqnarray}
where the unknown regression parameter $\beta$ is estimated by
    \begin{eqnarray} \label{ssec31:eq2}
      \hat{B}_r = \hat{G}_r^{-1} \left( \frac{1}{N}\sum_{i \in s} \omega_i r_i v_i^{-1} z_i y_i \right) & \textrm{with} & \hat{G}_r = \frac{1}{N}\sum_{i \in s} \omega_i r_i \hat{\phi}_i v_i^{-1} z_i z_i^{\top},
    \end{eqnarray}
where $\omega_i$ denotes a so called imputation weight, and $\hat{\phi}_i$ is given in (\ref{estim_phi_i}). The $\eta_i^*$'s are independently generated, and $\eta_i^*$ is equal to $1$ with the probability $\hat{\phi}_i$, and is equal to $0$ otherwise. \\

\noindent There are several possible choices for the imputation weights $\omega_i$. Using a modeling of the response mechanism for the variable $y_i$, \citet{haz:cha:nam:14} propose to choose the imputation weights so that $\hat{t}_{yI}$ is a doubly robust estimator for $t_y$. This means that the imputed estimator is approximately unbiased for $t_y$ whether the imputation model or the non-response model is adequately specified. \citet{haz:cha:nam:14} also prove that the resulting imputed estimator is consistent for $t_y$ under either approach. \\

\noindent The random imputation mechanism in (\ref{imput:mecha:1}) has three drawbacks. Firstly, it leads to an additional imputation variance due to the $\eta_i^*$'s. To overcome this problem, \citet{haz:cha:nam:14} proposed a balanced version of their imputation mechanism that is presented in Section \ref{ssec32}. Secondly, the imputation mechanism in (\ref{imput:mecha:1}) does not lead to an approximately unbiased estimator of the distribution function, as will be illustrated in the simulation study conducted in Section \ref{sec5}. Finally, the consistency of the imputed estimator $\hat{t}_{yI}$ relies on an assumption of mean square consistency for $\hat{B}_r$, which may be difficult to prove since the matrix $\hat{G}_r$ can be close to similarity for some samples. Following \citet{car:gog:lar:13} and \citet{cha:dop:18}, we introduce in Sections \ref{ssec33} and \ref{ssec34} a regularized version of $\hat{B}_r$.

\subsection{Haziza-Nambeu-Chauvet balanced imputation} \label{ssec32}

\noindent The balanced random imputation procedure of \citet{haz:cha:nam:14} consists in replacing a missing value with
    \begin{eqnarray} \label{imput:mecha:2}
      y_i^* & = & \tilde{\eta}_i^* \left\{z_i^{\top} \hat{B}_r \right\},
    \end{eqnarray}
where the $\tilde{\eta}_i^*$'s are not independently generated, but so that the imputation variance of $\hat{t}_{yI}$ is approximately equal to zero. Indeed, the imputation variance of $\hat{t}_{yI}$ is eliminated if the $\tilde{\eta}_i^*$'s are generated so that
    \begin{eqnarray} \label{ssec32:eq2}
      \sum_{i \in s} d_i(1-r_i)(\tilde{\eta}_i^*-\hat{\phi}_i)(z_i^{\top} \hat{B}_r) & = & 0.
    \end{eqnarray}
\citet{haz:cha:nam:14} propose a procedure adapted from the Cube method \citep{dev:til:04,cha:til:06} which enables to generate the $\tilde{\eta}_i^*$'s so that (\ref{ssec32:eq2}) is satisfied, at least approximately. As a result, the imputation variance is eliminated or at least significantly reduced. \\

\noindent This imputation procedure is called balanced random $\phi$-regression ($BRR_\phi$) imputation by \citet{haz:cha:nam:14}. They prove that under the $BRR_\phi$ imputation, an appropriate choice for the imputation weights $\omega_i$ leads to a doubly robust estimator for $t_y$. Also, their empirical results indicate that it performs well in reducing the imputation variance. A drawback of the $BRR_\phi$ imputation mechanism is that it does not preserve the distribution function of the imputed variable, because it does not take into account the error terms $\epsilon_i$ in the imputation model (\ref{imput:model}). This is empirically illustrated in section \ref{sec6}. To overcome this problem, two new imputation procedures are proposed in Sections \ref{ssec33} and \ref{ssec34}.

\subsection{Proposed random imputation} \label{ssec33}

\noindent The random imputation procedure that we propose consists in mimicking as closely as possible the imputation model (\ref{imput:model}), by replacing some missing $y_i$ with the imputed value
    \begin{eqnarray} \label{imput:mecha:3}
      y_i^* & = & \eta_i^* \left\{z_i^{\top} \hat{B}_{ar} + \sqrt{v_i} \epsilon_i^* \right\},
    \end{eqnarray}
where $\hat{B}_{ar}$ is a regularized version of $\hat{B}_r$, and $\eta_i^*$ is a Bernoulli random variable as defined in (\ref{imput:mecha:1}). The $\epsilon_i^*$'s are selected independently and with replacement in the set of observed residuals
    \begin{eqnarray} \label{set:obs:resid}
      E_r = \left\{e_j~;~r_j=1 \textrm{ and } \eta_j=1 \right\} & \textrm{ where } & e_j = \frac{y_j-z_j^{\top} \hat{B}_{ar}}{\sqrt{v_j}},
    \end{eqnarray}
with $Pr(\epsilon_i^*=e_{j})=\tilde{\omega}_j$ for any $j \in s$ such that $r_j=1$ and $\eta_j=1$, where
    \begin{eqnarray} \label{pr:epsi:st}
      \tilde{\omega}_j & = & \frac{\omega_{j}}{\sum_{k \in s} \omega_{j} r_k \eta_k}.
    \end{eqnarray}
We note
    \begin{eqnarray} \label{ssec33:eq0}
      \bar{e}_r = \sum_{j \in s} \tilde{\omega}_j r_j \eta_j e_j & \textrm{and} & \sigma_{er}^2 = \sum_{j \in s} \tilde{\omega}_j r_j \eta_j (e_j-\bar{e}_r)^2.
    \end{eqnarray}

\noindent The regularized version of $\hat{B}_{r}$ is obtained by following the approach in \citet{car:gog:lar:13} and \citet{cha:dop:18}. We first write
    \begin{eqnarray} \label{ssec33:eq1a}
      \hat{G}_r & = & \sum_{j=1}^p \alpha_{jr} v_{jr} v_{jr}^{\top},
    \end{eqnarray}
with $\alpha_{jr} \geq \ldots \geq \alpha_{pr}$ the non-negative eigenvalues of $\hat{G}_r$, and where $v_{1r},\ldots,v_{pr}$ are the associated orthonormal vectors. For some given $a>0$, the regularized versions of $\hat{G}_r$ and $\hat{B}_r$ are
    \begin{eqnarray} \label{ssec33:eq1b}
      \hat{G}_{ar} = \sum_{j=1}^p \max(\alpha_{jr},a) v_{jr} v_{jr}^{\top} & \textrm{and} & \hat{B}_{ar} = \hat{G}_{ar}^{-1} \left( \frac{1}{N}\sum_{i \in s} \omega_i r_i v_i^{-1} z_i y_i \right).
    \end{eqnarray}
The regularization leads to a matrix $\hat{G}_{ar}$ which is always invertible, and such that $\|\hat{G}_{ar}^{-1}\| \leq a^{-1}$ with $\|\cdot\|$ the spectral norm. \\

\noindent We prove in Section \ref{sec4} that $\hat{B}_{ar}$ is a mean-square consistent estimator of $\beta$, and that under the proposed imputation procedure the imputed estimator of the total is mean-square consistent for $t_y$. Also, we prove that the imputed estimator $\hat{F}_{I}(t)$ is $L_1$-consistent for the population distribution function. However, this imputation procedure leads to an additional variability for $\hat{t}_{yI}$ due to the imputation variance. Therefore, a balanced version is proposed in Section \ref{ssec34}.

\subsection{Proposed balanced imputation} \label{ssec34}

\noindent The balanced procedure consists in replacing a missing value with
    \begin{eqnarray} \label{imput:mecha:4}
      y_i^* & = & \tilde{\eta}_i^* \left\{z_i^{\top} \hat{B}_{ar} + \sqrt{v_i} \tilde{\epsilon}_i^* \right\},
    \end{eqnarray}
where the $\tilde{\eta}_i^*$'s and the $\tilde{\epsilon}_i^*$'s are not independently generated, but so as to eliminate the imputation variance of $\hat{t}_{yI}$. A sufficient condition for this consists in generating the residuals $\tilde{\eta}_i^*$ and $\tilde{\epsilon}_i^*$ so that
    \begin{eqnarray}
      \sum_{i \in s} d_i(1-r_i)(\tilde{\eta}_i^*-\hat{\phi}_i)(z_i^{\top} \hat{B}_r^*) & = & 0, \label{ssec34:eq1} \\
      \sum_{i \in s} d_i(1-r_i)\tilde{\eta}_i^* \sqrt{v_i} \tilde{\epsilon}_i^* & = & 0. \label{ssec34:eq2}
    \end{eqnarray}

\noindent This is done in a two-step procedure: first, the $\tilde{\eta}_i^*$'s are generated by means of Algorithm 1 in \citet{haz:cha:nam:14}, so that (\ref{ssec34:eq1}) is approximately respected; then, the $\tilde{\epsilon}_i^*$'s are generated by using Algorithm 1 described in \citet{cha:dev:haz:11}, so that (\ref{ssec34:eq2}) is approximately respected. \\

\noindent Since the balancing equations (\ref{ssec34:eq1}) and (\ref{ssec34:eq2}) are usually only approximately respected, the imputation variance is not completely eliminated, but it may be significantly reduced: see the simulation study in Section \ref{sec6}. Though the balanced imputation procedure is expected to provide estimators with smaller variance, the asymptotic properties of these estimators are difficult to study due to intricate dependencies introduced in the imputation process. Extending the results in Section \ref{sec4} is a challenging problem for further theoretical research.

\section{Properties of the proposed methods} \label{sec4}

\noindent To study the asymptotic properties of the sampling designs and estimators, we use the asymptotic framework of \citet{isa:ful:82}. We suppose that the population $U$ belongs to a nested sequence $\{U_\tau\}$ of finite populations with increasing sizes $N_\tau$, and that the vector of values for the variable of interest $y_{U\tau}=(y_{1\tau},\ldots,y_{N\tau})^{\top}$ belongs to a nested sequence $\{y_{U\tau}\}$ with increasing sizes $N_\tau$. For simplicity, the index $\tau$ is omitted in what follows and all limits are computed when $\tau \to \infty$. \\

\noindent We consider the following regularity assumptions:
    \begin{itemize}
      \item[H1:] Some constants $C_1,C_2>0$ exist, s.t. $C_1 \leq N n^{-1} \pi_i \leq C_2$ for any $i \in U$.
      \item[H2:] Some constant $C_3$ exists, s.t. $\sup_{i \neq j \in U} \left(n \left|1-\frac{\pi_{ij}}{\pi_i\pi_j}\right|\right) \leq C_3$.
      \item[H3:] Some constants $C_4,C'_4>0$ exist, s.t. $C_4 \leq \min_{i \in U} p_i$ and $C'_4 \leq \min_{i \in U} \phi_i$.
      \item[H4:] Some constants $C_5,C_6>0$ exist, s.t. $C_5 \leq N^{-1} n \omega_i \leq C_6$ for any $i \in U$.
      \item[H5:] Some constants $C_7,C_8,C_9>0$ exist, s.t. $C_7 \leq v_i \leq C_8$ and $\|z_i\| \leq C_9$ for any $i \in U$. Also, the matrix
        \begin{eqnarray} \label{sec4:eq1}
          G & = & \frac{1}{N} \sum_{i \in U} \omega_i \pi_i p_i \phi_i v_i^{-1} z_i z_i^{\top}
        \end{eqnarray}
      is invertible, and the constant $a$ chosen is s.t. $\|G^{-1}\| \leq a^{-1}$.
      \item[H6:] We have $E\left(\|\hat{\gamma}_r - \gamma\|^2\right)=O(n^{-1})$.
      \item[H7:] Some constant $C_{11}$ exists, s.t. for any vector $\tilde{\gamma}$
        \begin{eqnarray*}
          \left|f(u_i,\tilde{\gamma})-f(u_i,\gamma)\right| & \leq & C_{11} \|\tilde{\gamma}-\gamma\| \textrm{ for all } i \in U.
        \end{eqnarray*}
  \end{itemize}

  \noindent It is assumed in (H1) that the inclusion probabilities do not differ much from that obtained under simple random sampling, so that no design weight dominates the other. It is assumed in (H2) that the units in the population are not far from being independently selected: this assumption is verified for stratified simple random sampling and rejective sampling \citep{haj:64}, for example. It is assumed in (H3) that the response probabilities are bounded away from $0$, i.e. there is no hard-core non-respondents, and that the probabilities of observing a null value are also bounded away from $0$, i.e. the variable of interest is not degenerate. The assumption (H4) is related to the imputation weights, and is similar to assumption (H1). The assumption (H5) is related to the imputation model, and is necessary to control the behaviour of the regularized estimator $\hat{B}_{ar}$; see \citet{car:gog:lar:13} and \citet{cha:dop:18}. It is assumed in (H6) that the estimator $\hat{\gamma}_r$ is $\sqrt{n}$ mean-square consistent for the parameter $\gamma$. This assumption is somewhat strong, but is needed to obtain the standard rate of convergence for the imputed estimator of the total. It is assumed in (H7) that $f(\cdot,\cdot)$ is Lipschitz-Continuous in its second component. The assumptions (H5) and (H6) are also considered in \citet{haz:cha:nam:14}.

  \begin{prop} \label{prop1}
  Suppose that the imputation model in (\ref{imput:model}) holds and that the assumptions (H1)-(H7) are satisfied. Then we have
    \begin{eqnarray}
      E\left\{\|\hat{B}_{ar}-\beta\|^2\right\} & = & O(n^{-1}). \label{prop1:eq1}
    \end{eqnarray}
  \end{prop}

  \begin{prop} \label{prop2}
  Suppose that the imputation model in (\ref{imput:model}) holds and that the assumptions (H1)-(H7) are satisfied. Then under the random imputation mechanism proposed in Section \ref{ssec33}, we have
    \begin{eqnarray} \label{prop2:eq1}
      E\left[\left\{N^{-1}(\hat{t}_{yI}-t_{y})\right\}^2\right] & = & O(n^{-1}).
    \end{eqnarray}
  \end{prop}

  \begin{prop} \label{prop3}
  Suppose that the imputation model in (\ref{imput:model}) holds and that the assumptions (H1)-(H7) are satisfied. Also, suppose that the distribution function $F_{\epsilon}$ is absolutely continuous. Then under the random imputation mechanism proposed in Section \ref{ssec33}, we have for any $t \in \mathbb{R}$
     \begin{eqnarray} \label{prop3:eq1}
       E \left[\left\{\hat{F}_I(t)-F_N(t)\right\}^2\right] & = & o(1).
     \end{eqnarray}
  \end{prop}

\section{Variance estimation} \label{sec5}

\noindent We now consider variance estimation for the imputed estimator of the total $\hat{t}_{yI}$, under the proposed imputation procedures. The variance estimators are adapted from a linearized variance estimator proposed by \citet[][Section 2]{kim:rao:09} for deterministic/random regression imputation. They are obtained under a variance decomposition which makes use of the reverse approach \citep{fay:96,sha:ste:99}. For simplicity, we suppose that the $\phi_i$'s are modeled according to a logistic regression model and that the unknown parameter $\beta$ is the solution of the weighted estimated equation
    \begin{eqnarray} \label{sec5:eq1}
      \sum_{i \in s} \omega_i r_i u_i \left\{\eta_i-f(u_i,\gamma)\right\} & = & 0,
    \end{eqnarray}
with $\textrm{logit} f(u_i,\gamma)=u_i^{\top} \gamma$.

\subsection{Balanced imputation procedure} \label{ssec51}

\noindent We first consider the balanced imputation procedure proposed in Section \ref{ssec34}. We do not need to account for the imputation variance, since it is approximately eliminated for the estimation of the total with the proposed imputation procedure.  By following the approach of Kim and Rao~(2009), we obtain after some algebra the two-term variance estimator
    \begin{eqnarray} \label{sec5:eq2}
      \hat{V}_{BMRR}(\hat{t}_{yI}) & = & \hat{V}_1(\hat{t}_{yI}) + \hat{V}_2(\hat{t}_{yI}),
    \end{eqnarray}
see equations (10) and (13) in \citet{kim:rao:09}. The first term in the right-hand side of (\ref{sec5:eq2}) is
    \begin{eqnarray} \label{sec5:eq3}
      \hat{V}_1(\hat{t}_{yI}) & = & \sum_{i,j \in s} \left(\frac{\pi_{ij}-\pi_i \pi_j}{\pi_{ij}}\right) \hat{\xi}_i \hat{\xi}_j, \nonumber \\
      \textrm{with } \quad \hat{\xi}_i & = & d_i(\hat{\phi}_i z_i^{\top} \hat{B}_{ar}) + r_i \left(d_i+\omega_i \hat{\phi}_i v_i^{-1} \hat{a}^{\top} z_i \right)\left(y_i-\hat{\phi}_i z_i^{\top} \hat{B}_{ar}\right) \nonumber \\
                                        & + & r_i \omega_i (\hat{b}-\hat{c})^{\top} u_i \left(\eta_i-\hat{\phi}_i \right),
    \end{eqnarray}
with
    \begin{eqnarray} \label{sec5:eq3a}
      \hat{a} & = & \left( \sum_{i \in s} r_i \omega_i \hat{\phi}_i v_i^{-1} z_i z_i^{\top} \right)^{-1} \sum_{i \in s} d_i (1-r_i) \hat{\phi}_i z_i, \nonumber \\
      \hat{b} & = & \left( \sum_{i \in s} r_i \omega_i \hat{\phi}_i (1-\hat{\phi}_i) u_i u_i^{\top} \right)^{-1} \sum_{i \in s} d_i (1-r_i) \hat{\phi}_i (1-\hat{\phi}_i) (z_i^{\top}\hat{B}_{ar}) u_i, \\
      \hat{c} & = & \left( \sum_{i \in s} r_i \omega_i \hat{\phi}_i (1-\hat{\phi}_i) u_i u_i^{\top} \right)^{-1} \sum_{i \in s} \omega_i r_i v_i^{-1} \hat{\phi}_i (1-\hat{\phi}_i) (z_i^{\top}\hat{a}) (z_i^{\top}\hat{B}_{ar}) u_i, \nonumber
    \end{eqnarray}
and with $\pi_{ij}$ the probability that units $i$ and $j$ are selected together in the sample. The second term in the right-hand side of (\ref{sec5:eq2}) is
    \begin{eqnarray} \label{sec5:eq4}
      \hat{V}_2(\hat{t}_{yI}) = \sum_{i \in s} r_i d_i \left\{(1+\omega_i \pi_i v_i^{-1} \hat{a}^{\top} z_i) (y_i-\hat{\phi}_i z_i^{\top} \hat{B}_{ar})
                                                                +\omega_i \pi_i (\hat{b}-\hat{c})^{\top} u_i (\eta_i-\hat{\phi}_i) \right\}^2.
    \end{eqnarray}
As underlined by \citet{kim:rao:09}, $\hat{V}_2(\hat{t}_{yI})$ is not sensitive to a mis-specification of the covariance structure in model (\ref{imput:model}).

\subsection{Random imputation procedure} \label{ssec52}

\noindent We now consider the random imputation procedure proposed in Section \ref{ssec33}. We need to account for the additional variance due to the imputation process. By following once again the approach in \citet[][Section 4.1]{kim:rao:09}, we obtain the variance estimator
    \begin{eqnarray} \label{sec5:eq5}
      \hat{V}_{MRR}(\hat{t}_{yI}) & = & \hat{V}_{BMRR}(\hat{t}_{yI}) + \hat{V}_3(\hat{t}_{yI}),
    \end{eqnarray}
where $\hat{V}_{BMRR}(\hat{t}_{yI})$ is given in equation (\ref{sec5:eq2}), and with
    \begin{eqnarray} \label{sec5:eq6}
      \hat{V}_3(\hat{t}_{yI}) = \sum_{i \in s} d_i^2 (1-r_i) (y_i^*-\hat{\phi}_i z_i^{\top} \hat{B}_{ar})^2,
    \end{eqnarray}
with $y_i^*$ the imputed value given in equation (\ref{imput:mecha:3}).

\section{Simulation study} \label{sec6}

\noindent To evaluate the performance of the proposed imputation methods, we implement a simulation study inspired by \citet{haz:cha:nam:14}. We generate nine finite populations of size $N=10,000$ with an interest variable $y$ and four auxiliary variable $z_1,\ldots,z_4$. The values of $z_i,~i=1,\ldots,4,$ are generated according to a Gamma distribution with shift parameter $2$ and scale parameter $5$. The values of $y$ are generated according to the following mixture model:
    \begin{eqnarray} \label{generatoDPRop}
    y_i & = & \eta_i(a_0 + a_1 z_{1i} + a_2 z_{2i} + a_3 z_{3i} + a_4 z_{4i} + \epsilon_{i}),
    \end{eqnarray}
where the $\epsilon_i$'s are generated according to a standard normal distribution with variance $\sigma^2$. We use $a_0=30$ and $a_1=a_2=a_3=a_4=0.7$. Also, we choose three different values of $\sigma^2$ so that the coefficient of determination $R^2$ equals $0.4$, $0.5$ or $0.6$ for the units $i$ such that $\eta_i=1$. \\

\noindent The $\eta_i$'s are generated according to a Bernoulli distribution with parameter $\phi_i$, and
    \begin{eqnarray} \label{eq:phik}
     \log\left(\frac{\phi_i}{1-\phi_i}\right) & = & b_0 + b_1 z_{1i} + b_2 z_{2i} + b_3 z_{3i} + b_4 z_{4i},
    \end{eqnarray}
and different values for the parameters $b_0,\ldots,b_4$, chosen so that the proportion of non-null values is approximately equal to $0.60$, $0.70$, or $0.80$. The three different proportion of non-null values, crossed with the three different levels for the $R^2$, lead to the nine finite populations. \\

\noindent In each population, we select $R=1,000$ samples by means of rejective sampling \citep{haj:64} of size $n=500$, with inclusion probabilities proportional to the variable $z_{1i}$. In each sample, we generate a response indicator $r_i$ for unit $i$ according to a Bernoulli distribution with parameter $p_i$ such that
    \begin{eqnarray}\label{eq:pk}
     \log\left(\frac{p_i}{1-p_i}\right) & = & c_0 + c_1 z_{1i} + c_2 z_{2i} + c_3 z_{3i} + c_4 z_{4i}.
    \end{eqnarray}
We use different values for the parameters $c_0,\ldots,c_4$, chosen so that the proportion of respondents is approximately equal to $0.30$, $0.50$ or $0.70$.

\subsection{Properties of point estimators} \label{ssec61}

\noindent In this Section, we are interested in estimating the total $t_y$, and the distribution function $F_{N}(t)$ with $t=t_{\alpha}$, the $\alpha$-th quantile. In this simulation study, we consider the values $\alpha= 0.50, 0.75$ and $0.90$. We compare four imputation methods to handle non-response: (i) random imputation ($RR_\phi$) proposed by \citet{haz:cha:nam:14}, and presented in Section \ref{ssec31}; (ii) balanced random imputation ($BRR_\phi$) proposed by \citet{haz:cha:nam:14}, and presented in Section \ref{ssec32}; (iii) proposed random imputation method ($MRR_\phi$), presented in Section \ref{ssec33}; (iv) proposed balanced random imputation method ($BMRR_\phi$), presented in Section \ref{ssec34}. For each of the four methods, we use imputation weights $\omega_i=1$, and the $\phi_i$'s and $p_i$'s are estimated by means of logistic regression modeling. In each sample, missing values are replaced by imputed values according to imputation methods (i) to (iv), and the imputed estimators $\hat{t}_{yI}$ and $\hat{F}_I(t_{\alpha})$ are computed. \\

\noindent As a measure of bias of an estimator $\hat{\theta}_I$ of a finite population parameter $\theta$, we compute the Monte Carlo percent relative bias
    \begin{eqnarray} \label{biaisRelatif}
    RB_{MC}(\hat{\theta}_I) & = & \frac{100}{R} \sum_{k=1}^R \frac{(\hat{\theta}_{I(k)} -\theta)}{\theta}, \label{biaisRelatif}
    \end{eqnarray}
where $\hat{\theta}_{I(k)}$ denotes the imputed estimator computed in the $k$-th sample. As a measure of relative efficiency for each imputation method, using $BMRR_\phi$ as a benchmark, we computed
    \begin{eqnarray*}
    RE_{MC}(\hat{\theta}_I)=\frac{MSE_{MC}(\hat {\theta}_{I})}{MSE_{MC}(\hat {\theta}_{BMRR_\phi})} & \textrm{with} & MSE_{MC}(\hat{\theta}_I) = \frac{1}{R} \sum_{k=1}^R(\hat{\theta}_{I(k)} -\theta)^2, \label{erreurQuadratique}
    \end{eqnarray*}
the Mean Square Error of $\hat{\theta}_I$ approximated by means of the $R$ simulations. We observed no qualitative difference according to the different response rates. For brevity, we therefore only present the simulation results with an average proportion of respondents of $0.50$. The simulation results for the two other response rates are given in the Supplementary Material. \\

\noindent We first consider the estimation of the total $t_y$, for which the simulation results are given in Table \ref{tab:total:05}. The four imputation methods lead to approximately unbiased estimators of the total, as expected. Turning to the relative efficiency (RE), we note that in all studied cases the balanced version of an imputation method outperforms its unbalanced version. Also, the two balanced imputation procedures exhibit similar efficiency, with $BRR_\phi$ performing slightly better. This is likely due to fact that the balancing equations (\ref{ssec34:eq1}) and (\ref{ssec34:eq2}) are not exactly respected due to the landing phase of the cube method \citep[see][]{dev:til:04}. \\

\begin{table}
\begin{center}
\begin{tabular}{|ll|c|c|c|c|c|c|c|c|}
\hline
&  & \multicolumn{2}{|c|}{$RR_{\phi}$} & \multicolumn{2}{|c|}{$BRR_{\phi}$} & \multicolumn{2}{|c|}{$MRR_{\phi}$} & \multicolumn{2}{|c|}{$BMRR_{\phi}$} \\ \hline
$R^2$ & $\overline{\phi}$ & RB \% & RE & RB \% & RE & RB \% & RE & RB \% & RE \\ \hline
0.4 & 0.6 & 0.23 & 1.16 & 0.25 & 0.99 & 0.16 & 1.21 & 0.28 & 1.00 \\
0.4 & 0.7 & 0.11 & 1.07 & 0.26 & 0.96 & 0.09 & 1.14 & 0.32 & 1.00 \\
0.4 & 0.8 & 0.35 & 1.06 & 0.41 & 0.98 & 0.34 & 1.14 & 0.42 & 1.00 \\
\hline
0.5 & 0.6 & 0.33 & 1.09 & 0.27 & 0.99 & 0.31 & 1.12 & 0.24 & 1.00 \\
0.5 & 0.7 & 0.26 & 1.16 & 0.35 & 0.99 & 0.23 & 1.23 & 0.31 & 1.00 \\
0.5 & 0.8 & 0.44 & 1.13 & 0.44 & 0.99 & 0.43 & 1.21 & 0.45 & 1.00 \\
\hline
0.6 & 0.6 & 0.33 & 1.17 & 0.37 & 0.99 & 0.32 & 1.21 & 0.33 & 1.00 \\
0.6 & 0.7 & 0.18 & 1.13 & 0.35 & 0.99 & 0.15 & 1.18 & 0.35 & 1.00 \\
0.6 & 0.8 & 0.46 & 1.09 & 0.50 & 0.98 & 0.43 & 1.16 & 0.49 & 1.00 \\ \hline
\end{tabular}
\caption{Relative bias (RB \%) and Relative efficiency (RE) of four imputed estimators of the total with an average response probability of $50 \% $} \label{tab:total:05}
\end{center}
\end{table}

\noindent We now consider the estimation of the population distribution function, for which the simulation results are presented in Table \ref{tab:F50:F75:F90:05}. In all cases, the two proposed imputation methods $MRR_\phi$ and $BMRR_\phi$ lead to approximately unbiased estimators of the distribution function, with absolute relative biases no greater than 2 \% . On the contrary, the $RR_\phi$ and the $BRR_\phi$ imputation methods lead to biased estimators, and the absolute relative bias can be as large as 14 \% . We note that the bias is larger for the lower quantiles. Turning to the relative efficiency, we note that $MRR_\phi$ and $BMRR_\phi$ always outperform $RR_\phi$ and $BRR_\phi$, which is partly due to the bias under these latter imputation methods. Comparing the two proposed imputation methods, we note that $BMRR_\phi$ is systematically better than $MRR_\phi$ in terms of efficiency, with values of $RE$ ranging from $1.08$ to $1.32$ for $MRR_\phi$. \\

\begin{table}
\begin{center}
\begin{tabular}{|ll|c|c|c|c|c|c|c|c|}
\hline
&  & \multicolumn{2}{|c|}{$RR_{\phi}$} & \multicolumn{2}{|c|}{$BRR_{\phi}$} & \multicolumn{2}{|c|}{$MRR_{\phi}$} & \multicolumn{2}{|c|}{$BMRR_{\phi}$} \\ \hline
&  & RB \% & RE & RB \% & RE & RB \% & RE & RB \% & RE \\ \hline
$R^2$ & $\overline{\phi}$ & \multicolumn{8}{|c|}{$50 \% $ quartile} \\ \cline{1-10}
0.4 & 0.6 & -6.58  & 2.29 & -6.62  & 2.12 & 0.07  & 1.27 & -0.03 & 1.00 \\
0.4 & 0.7 & -12.03 & 4.37 & -12.23 & 4.36 & 0.90  & 1.24 & 0.76  & 1.00 \\
0.4 & 0.8 & -14.07 & 6.40 & -14.17 & 6.47 & 0.39  & 1.20 & 0.48  & 1.00 \\
0.5 & 0.6 & -6.78  & 2.22 & -6.71  & 2.11 & -0.07 & 1.17 & 0.00  & 1.00 \\
0.5 & 0.7 & -12.14 & 4.35 & -12.26 & 4.26 & 1.01  & 1.26 & 0.95  & 1.00 \\
0.5 & 0.8 & -12.97 & 6.38 & -12.94 & 6.24 & 0.64  & 1.32 & 0.79  & 1.00 \\
0.6 & 0.6 & -6.71  & 2.40 & -6.75  & 2.20 & 0.15  & 1.26 & 0.26  & 1.00 \\
0.6 & 0.7 & -12.06 & 4.58 & -12.26 & 4.59 & 0.98  & 1.23 & 0.66  & 1.00 \\
0.6 & 0.8 & -11.37 & 5.37 & -11.33 & 5.27 & 1.05  & 1.24 & 1.08  & 1.00 \\ \hline \hline
$R^2$ & $\overline{\phi}$ & \multicolumn{8}{|c|}{$75 \% $ quartile} \\ \cline{1-10}
0.4 & 0.6 & 6.80 & 4.13 & 6.83 & 4.15 & 1.45 & 1.18 & 1.31 & 1.00 \\
0.4 & 0.7 & 8.12 & 5.17 & 8.10 & 5.16 & 1.39 & 1.18 & 1.34 & 1.00 \\
0.4 & 0.8 & 8.07 & 5.64 & 8.06 & 5.63 & 0.60 & 1.23 & 0.72 & 1.00 \\
0.5 & 0.6 & 6.46 & 3.93 & 6.47 & 3.91 & 1.28 & 1.21 & 1.42 & 1.00 \\
0.5 & 0.7 & 7.61 & 4.81 & 7.61 & 4.81 & 1.27 & 1.18 & 1.34 & 1.00 \\
0.5 & 0.8 & 7.63 & 4.95 & 7.64 & 4.95 & 0.77 & 1.21 & 0.79 & 1.00 \\
0.6 & 0.6 & 6.12 & 3.68 & 6.12 & 3.64 & 1.39 & 1.21 & 1.52 & 1.00 \\
0.6 & 0.7 & 7.38 & 4.52 & 7.36 & 4.50 & 1.51 & 1.22 & 1.53 & 1.00 \\
0.6 & 0.8 & 7.14 & 4.49 & 7.15 & 4.48 & 0.80 & 1.17 & 0.86 & 1.00 \\ \hline \hline
$R^2$ & $\overline{\phi}$ & \multicolumn{8}{|c|}{$90 \% $ quartile} \\ \cline{1-10}
0.4 & 0.6 & 3.27 & 2.86 & 3.27 & 2.85 & 0.80 & 1.19 & 0.72 & 1.00 \\
0.4 & 0.7 & 3.55 & 2.89 & 3.55 & 2.89 & 0.98 & 1.08 & 0.91 & 1.00 \\
0.4 & 0.8 & 3.46 & 3.37 & 3.46 & 3.37 & 0.60 & 1.15 & 0.55 & 1.00 \\
0.5 & 0.6 & 3.10 & 2.64 & 3.10 & 2.63 & 0.74 & 1.21 & 0.79 & 1.00 \\
0.5 & 0.7 & 3.54 & 2.93 & 3.53 & 2.92 & 1.04 & 1.08 & 1.08 & 1.00 \\
0.5 & 0.8 & 3.43 & 3.43 & 3.43 & 3.43 & 0.69 & 1.18 & 0.64 & 1.00 \\
0.6 & 0.6 & 3.19 & 2.47 & 3.19 & 2.46 & 1.02 & 1.14 & 1.03 & 1.00 \\
0.6 & 0.7 & 3.39 & 2.87 & 3.39 & 2.86 & 1.08 & 1.16 & 1.06 & 1.00 \\
0.6 & 0.8 & 3.26 & 3.16 & 3.26 & 3.16 & 0.61 & 1.14 & 0.66 & 1.00 \\  \hline
\end{tabular}
\caption{Relative bias (RB \%) and Relative efficiency (RE) of four imputed estimators of the distribution function evaluated at the $50 \% $, $75 \% $ and $90 \% $ quartiles with an average response probability of $50 \% $} \label{tab:F50:F75:F90:05}
\end{center}
\end{table}

\noindent We also conducted additional simulations to evaluate the influence of the specific distribution used to simulate the random residuals in the imputation model. More precisely, we generated the variable of interest according to the mixture model presented in (\ref{generatoDPRop}), but with the residuals $\epsilon_i$'s generated either (a) from a gamma distribution or (b) from a log-normal distribution with variance $\sigma^2$. The simulation results are presented in the Supplementary Material. We observed no qualitative difference as compared to normally distributed residuals.

\subsection{Properties of variance estimators} \label{ssec62}

\noindent We now consider the properties of the variance estimators proposed in Section \ref{sec5}. Under the rejective sampling design used in the simulation study, we replace the component $\hat{V}_1(\hat{t}_{yI})$ given in (\ref{sec5:eq3}) with the Hajek-Rosen variance estimator
    \begin{eqnarray} \label{ssec62:eq1}
      \hat{V}_{HR,1}(\hat{t}_{yI}) = \frac{n}{n-1} \sum_{i \in s} (1-\pi_i) (\hat{\xi}_i-\hat{R})^2 \textrm{ with } \hat{R}=\frac{\sum_{i \in s} (1-\pi_i) \hat{\xi}_i}{\sum_{i \in s} (1-\pi_i)},
    \end{eqnarray}
see also \citet{cha:dop:18}. This leads to the simplified variance estimator
    \begin{eqnarray} \label{ssec62:eq2}
      \tilde{V}_{BMRR}(\hat{t}_{yI}) & = & \hat{V}_{HR,1}(\hat{t}_{yI}) + \hat{V}_2(\hat{t}_{yI}),
    \end{eqnarray}
for the proposed balanced imputation procedure $BMRR_{\phi}$, and to the simplified variance estimator
    \begin{eqnarray} \label{ssec62:eq3}
      \tilde{V}_{MRR}(\hat{t}_{yI}) & = & \tilde{V}_{BMRR}(\hat{t}_{yI}) + \hat{V}_3(\hat{t}_{yI}),
    \end{eqnarray}
for the proposed random imputation procedure $MRR_{\phi}$. \\

\noindent We computed the Monte-Carlo percent relative bias of these two variance estimators, using an independent simulation-based approximation of the true mean square error of $\hat{t}_{yI}$ based on $10,000$ simulations. We also computed the coverage rates of the associated normality-based confidence intervals, with nominal error rate of $2.5 \% $ in each tail. We only consider the two cases when the average proportion of respondents is $0.50$ and $0.70$. We first consider the results for $BMRR_{\phi}$, which are presented in Table \ref{evar:bal}. In all cases, the variance estimator $\tilde{V}_{BMRR}(\hat{t}_{yI})$ has a small bias, no greater than 6 \% . The variance estimator is slightly negatively biased with $\bar{p}=0.50$. This is likely due to the fact that the imputation variance is not completely eliminated with the proposed balanced imputation procedure, due to the landing phase of the cube method. The coverage rates are approximately respected in any case. We now turn to $MRR_{\phi}$, for which the simulation results are presented in Table \ref{evar:rand}. The variance estimator $\tilde{V}_{MRR}(\hat{t}_{yI})$ is approximately unbiased with $\bar{p}=0.50$, but is slightly positively biased with $\bar{p}=0.70$. The coverage rates are approximately respected in all cases.

\begin{table}
\begin{center}
\begin{tabular}{|l|cc|cc|cc|}
\hline
          & \multicolumn{6}{|c|}{Population 1} \\ \cline{2-7}
          & \multicolumn{2}{|c|}{$\bar{\phi}=0.6$} & \multicolumn{2}{|c|}{$\bar{\phi}=0.7$} & \multicolumn{2}{|c|}{$\bar{\phi}=0.8$} \\
          & $\bar{p}=0.5$ & $\bar{p}=0.7$ & $\bar{p}=0.5$ & $\bar{p}=0.7$ & $\bar{p}=0.5$ & $\bar{p}=0.7$ \\ \hline
RB (\% )  & -5.6 & 5.1  & -5.4 & 2.1  & -4.7 & 3.2 \\
Cov. Rate & 93.4 & 95.3 & 93.8 & 95.3 & 93.5 & 95.7 \\ \hline \hline
          & \multicolumn{6}{|c|}{Population 2} \\ \cline{2-7}
          & \multicolumn{2}{|c|}{$\bar{\phi}=0.6$} & \multicolumn{2}{|c|}{$\bar{\phi}=0.7$} & \multicolumn{2}{|c|}{$\bar{\phi}=0.8$} \\
          & $\bar{p}=0.5$ & $\bar{p}=0.7$ & $\bar{p}=0.5$ & $\bar{p}=0.7$ & $\bar{p}=0.5$ & $\bar{p}=0.7$ \\ \hline
RB (\% )  & -4.5 & 5.3  & -3.7 & 3.2  & -3.6 & 2.7 \\
Cov. Rate & 93.4 & 95.8 & 93.0 & 95.3 & 93.6 & 95.9 \\ \hline \hline
          & \multicolumn{6}{|c|}{Population 3} \\ \cline{2-7}
          & \multicolumn{2}{|c|}{$\bar{\phi}=0.6$} & \multicolumn{2}{|c|}{$\bar{\phi}=0.7$} & \multicolumn{2}{|c|}{$\bar{\phi}=0.8$} \\
          & $\bar{p}=0.5$ & $\bar{p}=0.7$ & $\bar{p}=0.5$ & $\bar{p}=0.7$ & $\bar{p}=0.5$ & $\bar{p}=0.7$ \\ \hline
RB (\% )  & -4.7 & 4.8  & -4.0 & 2.2  & -4.3 & 2.8 \\
Cov. Rate & 93.6 & 95.5 & 93.5 & 95.3 & 93.1 & 95.9 \\ \hline
\end{tabular}
\caption{Monte-Carlo percent relative bias of the variance estimator and coverage rate for the proposed balanced imputation procedure $BMRR_{\phi}$} \label{evar:bal}
\end{center}
\end{table}

\begin{table}
\begin{center}
\begin{tabular}{|l|cc|cc|cc|}
\hline
          & \multicolumn{6}{|c|}{Population 1} \\ \cline{2-7}
          & \multicolumn{2}{|c|}{$\bar{\phi}=0.6$} & \multicolumn{2}{|c|}{$\bar{\phi}=0.7$} & \multicolumn{2}{|c|}{$\bar{\phi}=0.8$} \\
          & $\bar{p}=0.5$ & $\bar{p}=0.7$ & $\bar{p}=0.5$ & $\bar{p}=0.7$ & $\bar{p}=0.5$ & $\bar{p}=0.7$ \\ \hline
RB (\% )  & -2.1 & 6.5  & -2.2 & 4.6  & -1.2 & 5.0 \\
Cov. Rate & 93.5 & 95.1 & 94.3 & 96.2 & 93.7 & 96.2 \\ \hline \hline
          & \multicolumn{6}{|c|}{Population 2} \\ \cline{2-7}
          & \multicolumn{2}{|c|}{$\bar{\phi}=0.6$} & \multicolumn{2}{|c|}{$\bar{\phi}=0.7$} & \multicolumn{2}{|c|}{$\bar{\phi}=0.8$} \\
          & $\bar{p}=0.5$ & $\bar{p}=0.7$ & $\bar{p}=0.5$ & $\bar{p}=0.7$ & $\bar{p}=0.5$ & $\bar{p}=0.7$ \\ \hline
RB (\% )  & -1.6 & 7.2  & -3.6 & 4.4  & -0.3 & 4.0 \\
Cov. Rate & 94.2 & 94.8 & 93.6 & 95.9 & 92.7 & 96.2 \\ \hline \hline
          & \multicolumn{6}{|c|}{Population 3} \\ \cline{2-7}
          & \multicolumn{2}{|c|}{$\bar{\phi}=0.6$} & \multicolumn{2}{|c|}{$\bar{\phi}=0.7$} & \multicolumn{2}{|c|}{$\bar{\phi}=0.8$} \\
          & $\bar{p}=0.5$ & $\bar{p}=0.7$ & $\bar{p}=0.5$ & $\bar{p}=0.7$ & $\bar{p}=0.5$ & $\bar{p}=0.7$ \\ \hline
RB (\% )  & -1.5 & 7.0  & -1.3 & 5.5  & 0.1  & 3.5 \\
Cov. Rate & 94.1 & 95.8 & 94.6 & 95.2 & 93.7 & 95.8 \\ \hline
\end{tabular}
\caption{Monte-Carlo percent relative bias of the variance estimator and coverage rate for the proposed random imputation procedure $MRR_{\phi}$} \label{evar:rand}
\end{center}
\end{table}

\section{Application} \label{sec:appli}

\noindent We apply the proposed imputation methods on data modelled from the Monthly Retail Trade Survey \citep{mul:oli:kap:14,boi:cha:haz:16,cha:dop:18}, which have been created to reproduce as closely as possible the original survey data. We consider the variable giving the sales ($y_{0i}$). We create in the dataset a domain indicator, equal to $1$ if the unit is in the domain and to $0$ otherwise. The variable of interest is $y_{i} = y_{0i} 1(i \in d)$, and we are interested in estimating the total and the distribution function of this variable. This case occurs when we are interested in domain estimation, and when the domain itself is not known for all sampled units due to non-response. For imputation purpose, we use as auxiliary variables a measure of size ($z_{1i}$), the prior month sales ($z_{2i}$) and the prior month inventories ($z_{3i}$). \\

\noindent The survey data arise from a stratified simple random sampling design with $6$ strata $U_h,~h=1,\ldots,6$. In this application, we leave apart the take-all stratum, which leads to five strata with sizes $N_h$ ranging from $463$ to $9~993$, and with sample sizes $n_h$ ranging from $57$ to $145$. The number of responding units per stratum $n_{rh}$ varies from $44$ to $75$. We suppose that the response mechanism is Missing At Random (MAR), and is explained by the strata indicators. In other words, we suppose that the response mechanism is uniform within each stratum. \\

\noindent The variable $y_{i}$ suffers from item non-response. We use an imputation model similar to that in \citet{boi:cha:haz:16}, but adapted to cover zero-inflated variables. More precisely, we suppose that each stratum $U_h$ is partitioned into $G_h$ imputation cells, obtained by ranking the units with respect to $z_{1i}$. The imputation model is
    \begin{eqnarray} \label{imput:model:appli}
      y_{i} & = & \eta_i \left\{\beta_{hg} + \epsilon_i \right\}
    \end{eqnarray}
for any unit $i$ belonging to the cell $g$ in stratum $U_h$. This is a particular case of the imputation model given in (\ref{imput:model:appli}), using for each stratum $U_h$ as auxiliary information $z_i$ the set of cell indicators. \\

\noindent We perform the imputation methods presented in Sections \ref{ssec33} and \ref{ssec34}, using equal imputation weights $\omega_i$. For any non-responding unit $i$, we obtain the estimated probability $\hat{\phi}_i$ through a logistic regression on the set $u_i=(1,z_{1i},z_{2i},z_{3i})^{\top}$ of auxiliary variables. Inside each stratum $U_h$, the estimator $\hat{B}_{arh}$ of $\beta_h$ is obtained from equation (\ref{ssec33:eq1b}), with $z_{i}$ the vector of cell indicators. We used $a=0.05$, and in this case no regularization was needed. The imputed values are then obtained from equation (\ref{imput:mecha:3}) for the proposed random imputation procedure, and from equation (\ref{imput:mecha:4}) for the proposed balanced random imputation procedure. \\

\noindent For each of the two imputation methods, we computed the imputed estimator of the total $\hat{t}_{yI}$ and the imputed estimator of the distribution function $\hat{F}_{I}(t)$ for several values of $t$. We also computed a with-replacement bootstrap variance estimator for the imputed estimators, see \cite{boi:cha:haz:16} and \cite{cha:dop:18}. The bootstrap is performed as if the samples were selected with replacement, which is reasonable in view of the small sampling rates inside strata. To compare the efficiency of the imputed estimators, we compute
    \begin{eqnarray} \label{appli:eq6}
      re & = & \frac{v_{boot}\{\hat{\theta}_{EBRI}\}}{v_{boot}\{\hat{\theta}_{BRI}\}}.
    \end{eqnarray}

\noindent The results are shown in Table \ref{tab:appli:results}. From the imputed data set, both imputation methods give similar results in terms of point estimation. Turning to relative efficiency, we note that the proposed exact balanced random imputation procedure yields more efficient estimations, with values of $re$ ranging from $0.87$ to $0.94$.

\renewcommand{\arraystretch}{1.0}
\begin{table}[htb!]
\begin{center}
\begin{tabular}{lcccccccccc} \hline
     & & $\hat{t}_{yI}$ & & \multicolumn{7}{c}{$\hat{F}_{yI}(t)$ with $t (\times 1,000)$} \\ \cline{5-11}
     & & $(\times 10^9)$  & & 300 & 700 & 1,000 & 2,000 & 5,000 & 8,000 & 10,000 \\ \hline \hline
EBRI & & $29.94$ & & $0.38$ & $0.51$ & $0.60$ & $0.75$ & $0.94$ & $0.98$ & $0.99$ \\
 BRI & & $30.44$ & & $0.37$ & $0.50$ & $0.60$ & $0.74$ & $0.94$ & $0.98$ & $0.99$ \\
re   & & $0.91$   & & $0.88$ & $0.88$ & $0.89$ & $0.87$ & $0.94$ & $0.92$ & $0.92$ \\ \hline
\end{tabular}
\caption{Imputed estimator of the total and of the distribution function, and estimated related efficiency with two imputation methods \label{tab:appli:results}}\end{center}
\end{table}

\section{Conclusion} \label{sec7}

\noindent In this paper, we considered imputation for zero-inflated data. We proposed two imputation methods which enable to respect the nature of the data, and which preserve the finite population distribution function. In particular, we proposed a balanced imputation method which enables to preserve the distribution of the imputed variable while being fully efficient for the estimation of a total. \\
\noindent Our imputation methods rely upon the mixture regression imputation model proposed by \citet{haz:cha:nam:14}. As mentioned by these authors, the proposed methods could be extended to more general mixture regression models, for example to handle count data. \\
\noindent In practice, we may not be interested in the distribution function in itself, but rather in complex parameters such as quantiles. Establishing the theoretical properties of estimators of such parameters under the proposed imputation procedures is a challenging task, and is currently under investigation.

\bibliographystyle{apalike}

\appendix

\section{Proof of Proposition \ref{prop1}} \label{app:pprop1}


  \begin{lemma} \label{lem1}
    We have $E\left\{\|\hat{G}_{r}-G\|^2\right\} = O(n^{-1})$.
  \end{lemma}

  \begin{lemproof}
  \noindent We can write $\hat{G}_{r}-G = \left(\hat{G}_{r}-\tilde{G}_{r}\right) + \left(\tilde{G}_{r}-G\right)$, where
    \begin{eqnarray} \label{plem1:eq1}
     \tilde{G}_{r} & = & \frac{1}{N}\sum_{i \in s} \omega_i r_i \phi_i v_i^{-1} z_i z_i^{\top}.
    \end{eqnarray}
  With a proof similar to that of Lemma 2 in \citet{cha:dop:18}, we obtain $E\left\{\|\tilde{G}_{r}-G\|^2\right\} = O(n^{-1})$. Also, we obtain from the assumptions:
    \begin{eqnarray} \label{plem1:eq2}
      \left\|\hat{G}_{r}-\tilde{G}_{r}\right\| & \leq & \frac{C_6 (C_9)^2 C_{11}}{C_7} \left\|\hat{\gamma}_r-\gamma\right\|,
    \end{eqnarray}
  so that the result follows from Assumption (H6).
  \end{lemproof}

  \noindent We can write $\hat{B}_{ar}-\beta=T_1-T_2+T_3$, where
    \begin{eqnarray}
      T_1 & = & \hat{G}_{ar}^{-1} \left\{\frac{1}{N} \sum_{i \in s} \omega_i r_i v_i^{-1} z_i (y_i - \phi_i z_i^{\top} \beta) \right\}, \nonumber \\
      T_2 & = & \hat{G}_{ar}^{-1} \left\{\frac{1}{N} \sum_{i \in s} \omega_i r_i v_i^{-1} (\hat{\phi}_i - \phi_i) z_i z_i^{\top} \right\}  \beta, \label{pprop1:eq1} \\
      T_3 & = & \hat{G}_{ar}^{-1} \left\{(\hat{G}_{r}-\hat{G}_{ar}) 1(\hat{G}_{ar} \neq \hat{G}_{r}) \right\} \beta. \nonumber
    \end{eqnarray}
  We have
    \begin{eqnarray} \label{pprop1:eq2}
      \|T_1\|^2 & \leq & \frac{a^{-2}}{N^2} \sum_{i,j \in S} r_i r_j \omega_i \omega_j v_i^{-1} v_j^{-1} z_i^{\top} z_j (y_i-\phi_i z_i^{\top} \beta) (y_j-z_j^{\top} \beta).
    \end{eqnarray}
  Since the sampling design is non-informative and the response mechanism is unconfounded, we can write $E(\|T_1\|^2) = E_{pq} E_m (\|T_1\|^2)$ and
    \begin{eqnarray} \label{pprop1:eq3}
      E(\|T_1\|^2) & \leq & E_{pq} \left[ \frac{a^{-2}}{N^2} \sum_{i \in s} r_i \omega_i^2 v_i^{-2} \left\{\sigma^2 \phi_i v_i + \phi_i (1-\phi_i) (z_i^{\top} \beta)^2 \right\} \right],
    \end{eqnarray}
  and from the assumptions we obtain $E(\|T_1\|^2)=O(n^{-1})$. Also, we have
    \begin{eqnarray} \label{pprop1:eq4}
      \left\|T_2\right\| & \leq & \frac{C_6 (C_9)^2 C_{11}}{a C_7} \left\|\hat{\gamma}_r-\gamma\right\|,
    \end{eqnarray}
  and from Assumption (H6) we obtain $E(\|T_2\|^2)=O(n^{-1})$. Finally, since $\|\hat{G}_{r}-\hat{G}_{ar}\|^2 \leq a^2$, we have
    \begin{eqnarray}
      E(\|T_3\|^2) & \leq & \|\beta\|^2 \times Pr(\hat{G}_{ar} \neq \hat{G}_{r}) \nonumber \\
                   & \leq & \frac{4 \|\beta\|^2}{(\alpha_p-a)^2} E\left\{\|\hat{G}_{r}-G\|^2\right\}, \label{pprop1:eq5}
    \end{eqnarray}
  where the second line in (\ref{pprop1:eq5}) follows from equation (B.21) in \citet{cha:dop:18}, and $\alpha_p$ is the largest eigenvalue of $G$ given in equation (\ref{sec4:eq1}). From Lemma \ref{lem1}, we have $E(\|T_3\|^2)=O(n^{-1})$, which completes the proof.

\section{Proof of Proposition \ref{prop2}} \label{app:pprop2}

\begin{lemma} \label{lem2}
We have
    \begin{eqnarray}
      E\left\{(\bar{e}_r)^2\right\} & = & O(n^{-1}), \label{lem2:eq1} \\
      E\left\{\sigma_{er}^2\right\} & = & O(1). \label{lem2:eq2}
    \end{eqnarray}
\end{lemma}

\begin{lemproof}

\noindent We consider equation (\ref{lem2:eq1}) only. The proof of equation (\ref{lem2:eq2}) is similar. We can rewrite $\bar{e}_r=T_4-T_5$, with
    \begin{eqnarray} \label{plem2:eq1}
      T_4 = \sum_{j \in s} \tilde{\omega}_j \eta_j r_j \epsilon_j & \textrm{and} & T_5 = \left(\sum_{j \in s} \tilde{\omega}_j \eta_j r_j v_j^{-1/2} z_j \right)^{\top} (\hat{B}_{ar}-\beta).
    \end{eqnarray}
It follows from the assumptions and from Proposition \ref{prop1} that $E(T_5^2)=O(n^{-1})$. \\

\noindent We can rewrite $E(T_4^2)=\sigma^2 E(T'_4)$, with $T'_4=\sum_{j \in s} \tilde{\omega}_j^2 \eta_j r_j$. We note $X = \sum_{j \in s} \omega_j r_j \eta_j$, and $m_X=\sum_{j \in s} \omega_j p_j \phi_j$. We can write $T'_4=T'_{41}+T'_{42}$, where $T'_{41}=T'_{4} 1(X > m_X/2)$ and $T'_{42}=T_{4} 1(X \leq m_X/2)$. From the assumptions, we have
    \begin{eqnarray}
      T'_{41} & \leq & \frac{4}{(C_4 C'_4 C_5)^2} \times \frac{1}{N^2} \sum_{i \in s} \omega_i^2 p_i \phi_i,
    \end{eqnarray}
which leads to $E(T'_{41})=o(n^{-1})$. Also, since $T'_4 \leq 1$, we have $T'_{42} \leq 1(X \leq m_X/2)$ and by using the Chebyshev inequality we obtain
    \begin{eqnarray}
      E(T'_{42}|s) & \leq & \frac{4}{(C_4 C'_4 C_5)^2} \times \frac{1}{N^2} \sum_{i \in s} \omega_i^2 (p_i \phi_i)(1-p_i \phi_i),
    \end{eqnarray}
which leads to $E(T'_{42})=o(n^{-1})$.
\end{lemproof}

\noindent From the assumptions, we have $E\left[\left\{N^{-1}(\hat{t}_{y\pi}-t_{y})\right\}^2\right] = O(n^{-1})$, so that it is sufficient to prove that $E\left[\left\{N^{-1}(\hat{t}_{yI}-\hat{t}_{y\pi})\right\}^2\right] = O(n^{-1})$. We have $N^{-1}(\hat{t}_{yI}-t_{y}) = T_6+T_7+T_8+T_9$, with
    \begin{eqnarray*}
      T_6 & = & N^{-1} \sum_{i \in s} d_i (1-r_i) (y_i^*-\hat{\phi}_i z_i^{\top} \hat{B}_{ar}), \label{T4} \\
      T_7 & = & N^{-1} \sum_{i \in s} d_i (1-r_i) \hat{\phi}_i z_i^{\top} (\hat{B}_{ar}-\beta), \label{T5} \\
      T_8 & = & N^{-1} \sum_{i \in s} d_i (1-r_i) (\hat{\phi}_i -\phi_{i}) z_i^{\top} \beta, \label{T6} \\
      T_9 & = & N^{-1} \sum_{i \in s} d_i (1-r_i) (\phi_{i} z_i^{\top} \beta-y_i). \label{T7}
    \end{eqnarray*}
It readily follows from the assumptions, equation (\ref{sec4:eq2}) and Proposition \ref{prop1}, that $E(T_7^2)=o(1)$ and $E(T_8^2)=o(1)$. Also, since $E_m(T_9)=0$, we obtain
    \begin{eqnarray*}
      E(T_9^2)=EV_m(T_9)=E \left[N^{-2} \sum_{i \in s} d_i^2 (1-r_i) \left\{\sigma^2 \phi_i v_i + \phi_i (1-\phi_i) (z_i^{\top} \beta)^2 \right\} \right],
    \end{eqnarray*}
which is $O(n^{-1})$. Therefore, we only need to focus on $T_6$, for which we have
    \begin{eqnarray*}
      E_I(T_6^2) & = & \left\{N^{-1} \sum_{i \in s} d_i(1-r_i) \hat{\phi}_i \sqrt{v_i} \right\}^2 (\bar{e}_r)^2 \\
               & + & N^{-2} \sum_{i \in s} d_i^2(1-r_i) \left\{\hat{\phi}_i(1-\hat{\phi}_i)(z_i^{\top} \hat{B}_{ar}+\sqrt{v_i} \bar{e}_r)^2 + \hat{\phi}_i v_i \sigma_{er}^2 \right\}. \nonumber
    \end{eqnarray*}
From Proposition \ref{prop1} and Lemma \ref{lem2}, we obtain $E(T_6^2)=O(n^{-1})$.

\section{Proof of Proposition \ref{prop3}}

\noindent From the assumptions, we have $E\left[\left\{\hat{F}_N(t)-F_N(t)\right\}^2\right] = O(n^{-1})$, so that it is sufficient to prove that $E\left[\left\{\hat{F}_I(t)-\hat{F}_N(t)\right\}^2\right] = o(1)$. We have $hat{F}_I(t)-\hat{F}_N(t)=T_{10}+T_{11}+T_{12}$, where
    \begin{eqnarray}
      T_{10} & = & N^{-1} \sum_{i \in s} d_i(1-r_i)\left\{1(y_i^* \leq t)-1(y_i^{**} \leq t)\right\}, \label{term5} \\
      T_{11} & = & N^{-1} \sum_{i \in s} d_i(1-r_i)\left\{1(y_i^{**} \leq t)-1(\hat{y}_i \leq t)\right\}, \label{term6} \\
      T_{12} & = & N^{-1} \sum_{i \in s} d_i(1-r_i)\left\{1(\hat{y}_i \leq t)-1(y_i \leq t)\right\}. \label{term7}
    \end{eqnarray}
The values $y_i^{**}$ and $\hat{y}_i$ are obtained as follows. We take
    \begin{eqnarray} \label{yi:hat}
      \hat{y}_i & = & \eta_i \left\{z_i^{\top} \beta + \sqrt{v_i} \hat{\epsilon}_i \right\},
    \end{eqnarray}
where $\hat{\epsilon}_i$ is selected with-replacement from the set $E'_r = \left\{\epsilon_j ~;~r_j=1 \textrm{ and } \eta_j=1 \right\}$. We note $j(i)$ the donor selected for unit $i$, so that $\hat{\epsilon}_i=\epsilon_{j(i)}$. Also, we take
    \begin{eqnarray} \label{yi:starstar}
      y_i^{**} & = & \eta_i \left\{z_i^{\top} \hat{B}_{ar} + \sqrt{v_i} e_{g(i)} \right\}
                 =   \eta_i \left\{z_i^{\top} \hat{B}_{ar} + \sqrt{v_i} \epsilon_{i}^* \right\}.
    \end{eqnarray}

\noindent We consider the term $T_{10}$ first. We can write
    \begin{eqnarray} \label{app3:eq1}
    1(y_i^* \leq t) - 1(y_i^{**} \leq t) & = & (\eta_i^*-\eta_i)\{1(\varepsilon_i^* \leq \hat{t}_i )- 1(t\geq 0) \},
    \end{eqnarray}
with $\hat{t}_i = v_i^{-1/2} (t-z_i^{\top} \hat{B}_{ar}$. This leads to $(T_{10})^2=T_{10,1}+T_{10,2}$, with
    \begin{eqnarray*}
      T_{10,1} & = & N^{-2} \sum_{i \in s} d_i^2(1-r_i)(\eta_i^*-\eta _i)^2\{1(\varepsilon_i^* \leq \hat{t}_i )- 1(t\geq 0) \}^2, \label{app3:eq2} \\
      T_{10,2} & = & N^{-2} \sum_{i\neq j \in s} d_i(1-r_i)d_j(1-r_j)(\eta_i^*-\eta_i)(\eta_j^*-\eta _j) \times \nonumber \\
             &   & \phantom{N^{-2} \sum_{i\neq j \in s} d_i(1-r_i)} \{1(\varepsilon_i^* \leq \hat{t}_i )- 1(t\geq 0)\}
      \{1(\varepsilon_j^* \leq \hat{t}_j )- 1(t\geq 0)\}. \label{app3:eq3}
    \end{eqnarray*}
From the assumptions, $T_{10,1}=O(n^{-1})$. Also, since $\eta_i^*$, $\eta_j^*$, $\varepsilon_i^*$ and $\varepsilon_j^*$ are independent with respect to the imputation mechanism, we obtain successively
    \begin{eqnarray*} \label{app3:eq5}
    E_I(T_{10,2}) & = & N^{-2} \sum_{i\neq j \in s} d_i(1-r_i)d_j(1-r_j)(\hat{\phi}_i-\eta _i)(\hat{\phi}_j-\eta _j) \times \nonumber \\
                &   & \phantom{N^{-2} \sum_{i\neq j \in s} d_i(1-r_i)}\{\hat{F}_{\varepsilon_r}(\hat{t}_i )- 1(t\geq 0)\}\{\hat{F}_{\varepsilon_r}(\hat{t}_j )- 1(t\geq 0)\}, \\
    E_m\{E_I(T_{10,2})| \varepsilon_j,j \in s; \eta _g, g \in S_r \} & = &
    N^{-2} \sum_{i\neq j \in s} d_i(1-r_i)d_j(1-r_j)(\hat{\phi}_i-\phi _i)(\hat{\phi}_j-\phi _j) \times \nonumber \\
    & & \phantom{\sum_{i\neq j \in s}} \{\hat{F}_{\varepsilon_r}(\hat{t}_i )- 1(t\geq 0)\}\{\hat{F}_{\varepsilon_r}(\hat{t}_j )- 1(t\geq 0)\},
    \end{eqnarray*}
where $\hat{F}_{\varepsilon_r}(t)=\sum_{j \in s} \tilde{\omega}_j r_j \eta_j 1(e_j\leq t)$. This leads to
    \begin{eqnarray*} \label{app3:eq7}
    E(T_{10,2}) & \leq & \left(\frac{C_{11}}{C_1}\right)^2 E\left(\|\hat{\gamma}_r-\gamma\|^2 \right) = o(1).
    \end{eqnarray*}
Consequently, $E(T_{10}^2)=o(1)$. \\

\noindent We now consider $T_{11}$, that we can write as
    \begin{eqnarray*} \label{app3:eq9}
    T_{11} & = & N^{-1} \sum_{i \in s} d_i(1-r_i)\eta_i\{1(\varepsilon_i^* \leq \hat{t}_i )- 1(\hat{\varepsilon}_i\leq t_i) \}
    \end{eqnarray*}
with $t_i = v_i^{-1/2} (t-z_i^{\top} \beta$, which leads to
    \begin{eqnarray*} \label{app3:eq10}
    E_I(|T_{11}|) & \leq & N^{-1} \sum_{i \in s} d_i(1-r_i)\eta_i \sum_{j \in s} \tilde{\omega}_j r_j \eta_j \mid 1(e_j \leq \hat{t}_i )- 1(\varepsilon_j\leq t_i) \mid \\
                  & \leq  & N^{-1} \sum_{i \in s} d_i(1-r_i) \eta_i \sum_{j \in s} \tilde{\omega}_j r_j \eta_j \mid 1(\varepsilon_j \leq t_{ij}) - 1(\varepsilon_j \leq t_i)\mid \equiv T'_{11},
    \end{eqnarray*}
with
    \begin{eqnarray*}
    t_{ij} & = & t_i + \left(\frac{z_j}{\sqrt{v_j}}-\frac{z_i}{\sqrt{v_i}}\right)^{\top}(\hat{B}_{ar}-\beta).
    \end{eqnarray*}

\noindent Let us take some constant $\nu >0$. Since the distribution function $F_\varepsilon$ is absolutely continuous, there exists some $\tau_\nu$ such that
    \begin{eqnarray*} \label{app3:eq10b}
      |t-u|\leq \tau_\nu & \Rightarrow & | F_\varepsilon(t)- F_\varepsilon(u) | \leq \nu
    \end{eqnarray*}
We note $1_A = 1\left(\|\hat{B}_{ar}-\beta\| \geq 0.25 \tau_\nu \sqrt{C_7}/C_9 \right)$, and $1_B=1-1_A$. We have $E\{T'_{11} 1(A)\} \leq (C_1)^{-1} E\{1(A)\}$, which is $o(1)$ from Proposition \ref{prop1} and the Chebyshev inequality. Also, we have
    \begin{eqnarray*}
      T'_{11} 1(B) & \leq & N^{-1} \sum_{i \in s} d_i(1-r_i) \eta_i \sum_{j \in s} \tilde{\omega}_j r_j \eta_j 1\left(t_i - \frac{\tau_\nu}{2}  \leq \varepsilon_j \leq t_i + \frac{\tau_\nu}{2}\right).
    \end{eqnarray*}
This leads to $E_m\{T'_{11} 1(B)\} \leq (C_1)^{-1} \nu$, and since $\nu$ is arbitrary small, $E\{T'_{11} 1(B)\} = o(1)$. Consequently, $E(|T_{11}|)=o(1)$. \\

\noindent Finally, we now consider $T_{12}$ that we can write as
    \begin{eqnarray*}
        T_{12}& = & N^{-1} \sum_{i \in s} d_i(1-r_i) \eta_i \left\{1(\hat{\varepsilon}_i \leq t_i )- 1(\varepsilon_i\leq t_i)\right\}. \nonumber
    \end{eqnarray*}
This successively leads to
    \begin{eqnarray}
        T_{12}& = & N^{-1} \sum_{i \in s} d_i(1-r_i) \eta_i \left\{1(\hat{\varepsilon}_i \leq t_i )- 1(\varepsilon_i\leq t_i)\right\}, \label{EmEI:T7} \\
        E_I(T_{12}) & = & N^{-1} \sum_{i \in s} d_i(1-r_i) \eta_i \sum_{j \in s}\tilde{\omega}_j r_j \eta_j \left\{1(\varepsilon_j \leq t_i )- 1(\varepsilon_i\leq t_i)\right\}, \nonumber \\
        E_m\{E_I(T_{12})|\eta_i,i\in s \}&=&N^{-1} \sum_{i \in s} d_i(1-r_i) \eta_i \sum_{j \in s}\tilde{\omega}_j r_j \eta_j \left\{F_\varepsilon( t_i )- 1(F_\varepsilon( t_i )\right\} =0, \nonumber
    \end{eqnarray}
and $E(T_{12})=0$, which gives
    \begin{eqnarray} \label{app3:eq16}
      E\{(T_{12})^2\} & = & E_p E_q E_m V_I(T_{12}) + E_p E_q V_m E_I(T_{12}) .
    \end{eqnarray}
We have $V_I(T_{12}) \leq C_1^{-1} n^{-1}$, so that the first term in the r.h.s. of (\ref{app3:eq16}) is $O(n^{-1})$. From the third line in equation (\ref{EmEI:T7}), we obtain
    \begin{eqnarray}
    V_m\{E_I(T_{12})\} & = & E_m V_m \{E_I(T_{12})|\eta_i,i\in s\}, \label{VmEI:T7}
    \end{eqnarray}
and from the rewriting
    \begin{eqnarray*} \label{app3:eq17}
      E_I(T_{12}) = N^{-1}\sum_{j \in s}\tilde{\omega}_j r_j \eta_j \sum_{i \in s} d_i(1-r_i) \eta_i  1(\varepsilon_j \leq t_i )
               - N^{-1} \sum_{i \in s} d_i(1-r_i) \eta_i 1(\varepsilon_i \leq t_i),
    \end{eqnarray*}
we obtain
    \begin{eqnarray} \label{app3:eq18}
    V_m\{E_I(T_{12})|\eta_i,i\in s\}
    & = & N^{-2}\sum_{j \in s} \tilde{\omega}_j^2 r_j \eta_j V_m\{ \sum_{i \in s} d_i(1-r_i) \eta_i 1(\varepsilon_j \leq \ t_i )|\eta_i,i\in s\}\nonumber \\
    & + & N^{-2} \sum_{i \in s} d_i^2(1-r_i) \eta_i F_\varepsilon(t_i) \{1-F_\varepsilon(t_i)\} \nonumber \\
    & = & N^{-2}(\sum_{i \in s} d_i)^2\sum_{j \in s}\tilde{\omega}_j^2 r_j \eta_j V_m\left\{\left. \frac{\sum_{i \in s} d_i(1-r_i) \eta_i 1(\varepsilon_j \leq \ t_i )}{\sum_{i \in s} d_i} \right| \eta_i,i\in s\right\} \nonumber \\
    & + & N^{-2} \sum_{i \in s} d_i^2(1-r_i) \eta_i F_\varepsilon(t_i) \{1-F_\varepsilon(t_i)\} \nonumber \\
    & \leq & N^{-2}(\sum_{i \in s} d_i)^2 \sum_{j \in s} \tilde{\omega}_j^2 \eta_j r_j + N^{-2} \sum_{i \in s} d_i^2 \nonumber. \\
    & \leq & \frac{\sum_{j \in s} \tilde{\omega}_j^2 \eta_j r_j + n^{-1}}{C_1^2}.
    \end{eqnarray}
From the proof of Lemma \ref{lem2}, we have $E(\sum_{j \in s} \tilde{\omega}_j^2 \eta_j r_j)=O(n^{-1})$. From (\ref{VmEI:T7}) and (\ref{app3:eq18}), we obtain that the second term in the r.h.s. of (\ref{app3:eq16}) is $O(n^{-1})$. Consequently, $E(T_{12}^2)=O(n^{-1})$. This completes the proof.

\end{document}